\documentclass[acmtocl]{acmtrans2m}

\acmVolume{TBD}
\acmNumber{TBD}
\acmYear{TBD}
\acmMonth{TBD} 

\newtheorem{theorem}{Theorem}[section]

\newtheorem{lemma}[theorem]{Lemma}
\newdef{definition}[theorem]{Definition}
\newdef{remark}[theorem]{Remark}
\newtheorem{example}{Example}[section] 

\title{A Dynamic Approach to Characterizing 
Termination of General Logic Programs}
\author{YI-DONG SHEN\\Chongqing University\\
JIA-HUAI YOU, LI-YAN YUAN and SAMUEL S. P. SHEN\\University of Alberta
\and
QIANG YANG\\Simon Fraser University}

\begin{abstract}
We present a new characterization of termination of general
logic programs. Most existing termination analysis approaches 
rely on some static information about the structure 
of the source code of a logic program, such as  
modes/types, norms/level mappings, models/interargument 
relations, and the like. We propose a dynamic approach 
which employs some 
key dynamic features of an infinite (generalized) SLDNF-derivation,
such as repetition of selected subgoals and recursive 
increase in term size. We also introduce a new formulation of
SLDNF-trees, called generalized SLDNF-trees. 
Generalized SLDNF-trees deal with
negative subgoals in the same way as Prolog and 
exist for any general logic programs.
\end{abstract}






\begin{document}

\begin{bottomstuff}
Author's address: Yi-Dong Shen, Department of Computer Science, 
Chongqing University, Chongqing 400044, China. 
Email: ydshen@cs.ualberta.ca.
Jia-Huai You and Li-Yan Yuan, Department of Computing Science, University of 
Alberta, Edmonton, Canada T6G 2H1. Email: \{you, yuan\}@cs.ualberta.ca.
Samuel S. P. Shen, Department of Mathematics, University of 
Alberta, Edmonton, Canada T6G 2H1. Email: shen@ualberta.ca.
Qiang Yang, School of Computing Science, 
Simon Fraser University, Burnaby, BC, Canada V5A 1S6.
Email: qyang@cs.sfu.ca.
\end{bottomstuff}
\maketitle 

\section{Introduction} 
For a program in any computer language, in addition to having to be logically 
correct, it should be terminating. Due to frequent use of recursion in 
logic programming, however, a logic program may more likely be 
non-terminating than a procedural program. Termination of 
logic programs then becomes an important topic in 
logic programming research. Because the problem is extremely hard (undecidable 
in general), it has been considered as a {\em never-ending story}; see 
\cite{DD93} for a comprehensive survey. 
 
The goal of termination analysis is to establish a characterization 
of termination of a logic program and design algorithms for automatic 
verification. A lot of methods for termination analysis have been 
proposed in the last decade. 
A majority of these existing methods are the {\em norm-} or {\em level  
mapping-based} approaches, which consist of inferring mode/type information, inferring 
norms/level mappings, inferring models/interargument relations, and verifying 
some well-founded conditions (constraints). For example, 
Ullman and Van Gelder \cite{UVG88} and  
Pl\"{u}mer \cite{Pl90a,Pl90b} focused on  
establishing a decrease in term size of some recursive calls based on 
interargument relations; Apt, Bezem and Pedreschi \cite{Apt1,Bezem92}, 
and Bossi, Cocco and Fabris \cite{BCF94} 
provided characterizations of Prolog left-termination based on level 
mappings/norms and models; Verschaetse \cite{V92}, Decorte, De Schreye 
and Fabris \cite{DDF93}, and Martin, King and Soper \cite{MKS96} exploited inferring 
norms/level mappings from mode and type information;   
De Schreye and Verschaetse \cite{DV95}, Brodsky and Sagiv \cite{BS91}, 
and Lindenstrauss and Sagiv \cite{LS97} discussed 
automatic inference of interargument/size relations;  
De Schreye, Verschaetse and Bruynooghe \cite{DVB92} 
addressed automatic verification of the well-founded constraints. 
Very recently, Decorte, De Schreye and Vandecasteele \cite{DDV99} 
presented an elegant unified termination analysis that integrates all 
the above components to produce a set of constraints that, when solvable,  
yields a termination proof. 
 
It is easy to see that the above methods  
are {\em compile-time} (or {\em static}) 
approaches in the sense that they make termination 
analysis only relying on some {\em static} information about the structure 
(of the source code) of a logic program, such as  
modes/types, norms (i.e. term sizes of atoms of clauses)/level  
mappings, models/interargument relations, and the like. 
Our observation shows that some {\em dynamic} 
information about the structure of a concrete infinite SLDNF-derivation, 
such as {\em repetition} of selected subgoals and {\em recursive 
increase} in term size, plays a crucial role in characterizing the  
termination. Such dynamic features are hard to capture by
applying a compile-time approach. This suggests that methods 
of extracting and utilizing  
dynamic features for termination analysis are worth exploiting.  
 
In this note, we present a {\em dynamic} approach 
by employing dynamic features 
of an infinite (generalized) SLDNF-derivation to
characterize termination of general logic programs. 
In Section 2, we introduce a notion 
of a generalized SLDNF-tree, which is the basis of our method. Roughly 
speaking, a generalized SLDNF-tree is a set of standard 
SLDNF-trees augmented  
with an ancestor-descendant relation on their subgoals. In Section 3, we  
define a key concept, {\em loop goals}, which captures both repetition 
of selected subgoals and recursive increase in term size of these 
subgoals. We then prove 
a necessary and sufficient condition for an infinite generalized 
SLDNF-derivation in terms of loop goals. 
This condition allows us to establish a dynamic
characterization of termination of general logic programs (Section 4).
In Section 5, we mention the related work, and
in Section 6 we conclude the paper with our future work. 
 
\subsection{Preliminaries}  
 
We present our notation and review some standard  
terminology of logic programs as described in \cite{Ld87}. 
 
Variables begin with a capital letter, and predicate, function  
and constant symbols with a lower case letter.  
A term is a constant, a variable, or a function
of the form $f(T_1, ..., T_m)$ where $f$ is a function symbol 
and the $T_i$s are terms. 
An atom is of the form 
$p(T_1, ..., T_m)$ where $p$ is a predicate symbol 
and the $T_i$s are terms. A literal is of the form 
$A$ or $\neg A$ where $A$ is an atom. Let $A$ be an atom/term. 
The size of $A$, denoted $|A|$, is the 
number of occurrences of function symbols,  
variables and constants in $A$.
By $\{A_i\}_{i=1}^n$ we denote a sequence $A_1$, $A_2$, ..., $A_n$. 
Two atoms, $A$ and $B$, are said to be {\em variants} if  
they are the same up to variable renaming.  
 
Lists are commonly used terms. A list is of the form 
$[]$ or $[T|L]$ where $T$ is a term and $L$ is a list. 
For our purpose, the symbols $[$, $]$ and $|$ in a list  
are treated as function symbols. 
 
\begin{definition} 
A (general) {\em logic program} is a finite set 
of clauses of the form 
 
$\qquad A\leftarrow L_1,..., L_n$ 
 
\noindent where $A$ is an atom and $L_i$s are literals.  
When $n=0$, the ``$\leftarrow$'' symbol is omitted. 
$A$ is called the {\em head} and $L_1,...,L_n$ is called the 
{\em body} of the clause. If a general logic program has no clause with negative 
literals like $\neg A$ in its body, it is called a {\em positive} logic program. 
\end{definition} 
 
\begin{definition}  
A {\em goal} is a headless clause 
$\leftarrow L_1,..., L_n$ where each literal $L_i$ is called a {\em subgoal}. 
$L_1,..., L_n$ is called a (concrete) {\em query}.  
\end{definition} 
 
The initial goal, $G_0=\leftarrow L_1,..., L_n$, is called 
a {\em top} goal. Without loss of generality, we shall assume throughout 
the paper that a top goal consists only of one atom, i.e. $n=1$ and $L_1$ 
is a positive literal.  
 
\begin{definition} 
A {\em control strategy} consists of two rules, one for selecting  
a goal from among a set of goals and the other for selecting 
a subgoal from the selected goal.  
\end{definition} 
 
The second rule in a control strategy is usually called  
a {\em selection} or {\em computation} 
rule in the literature. To facilitate our presentation,
throughout the paper we choose to use the best-known 
{\em depth-first, left-most} control strategy (used in Prolog)
to describe our approach (It can be adapted
to any other fixed control strategies).  
So the {\em selected} subgoal in each goal is the 
left-most subgoal. Moreover, the clauses in a logic program 
are used in their textual order. 
 
Trees are commonly used to represent  
the search space of a top-down proof procedure. For convenience, 
a node in such a tree is represented by $N_i:G_i$ where $N_i$ is the name 
of the node and $G_i$ is a goal labeling the node. Assume no two  
nodes have the same name. Therefore, we can refer to nodes by their names. 
 
\section{Generalized SLDNF-Trees} 
 
In order to characterize infinite 
derivations more precisely, 
in this section we extend the standard SLDNF-trees  
\cite{Ld87} to include some new features.  
We first define the ancestor-descendant relation on selected subgoals. 
Informally, $A$ is an ancestor subgoal of $B$ if the proof of $A$ 
needs (or in other words goes via) the proof of $B$. For example, 
let $M:\leftarrow A,A_1,...,A_m$ be a node in an SLDNF-tree, and 
$N:\leftarrow B_1,...,B_n,A_1,...,A_m$ be a child node of $M$ that 
is generated by resolving $M$ on the subgoal $A$ with a clause  
$A\leftarrow B_1,...,B_n$. Then $A$ at $M$ is an ancestor subgoal 
of all $B_i$s at $N$. However, such relationship does not 
exist between $A$ at $M$ and any $A_j$ at $N$. It is easily seen 
that all $B_i$s at $N$ inherit the ancestor subgoals of $A$ at $M$. 
 
The ancestor-descendant relation can be explicitly expressed using 
an {\em ancestor list}. 
The ancestor list of a subgoal $A$ at a node $M$, 
denoted $AL_{A@M}$, is of the form $\{(N_1,D_1),...,(N_l,D_l)\}$ $(l\geq 0)$, 
where for each $(N_i, D_i)\in AL_{A@M}$, $N_i$ is a node 
name and $D_i$ a subgoal such that $D_i$ at $N_i$
is an ancestor subgoal of $A$ at $M$.
For instance, in the above example,  
the ancestor list of each $B_i$ at node $N$ 
is $AL_{B_i@N}=\{(M,A)\}\cup AL_{A@M}$ and
the ancestor list of each $A_i$ at node $N$ 
is $AL_{A_i@N}=AL_{A_i@M}$. 

Let $N_i:G_i$ and $N_k:G_k$ be two nodes and 
$A$ and $B$ be the selected subgoals in $G_i$ and $G_k$, 
respectively. We use  $A\prec_{anc}B$  to denote that
$A$ is an ancestor subgoal of $B$.
When $A$ is an ancestor subgoal of $B$, we refer to $B$ as a {\em descendant 
subgoal} of $A$, $N_i$ as an {\em ancestor node} of $N_k$, and $N_k$ as  
a {\em descendant node} of $N_i$.  

Augmenting SLDNF-trees with ancestor lists leads to the  
following definition of SLDNF$^*$-trees. 
 
\begin{definition}[(SLDNF$^*$-trees)]
\label{tree} 
Let $P$ be a general logic program,
$G_r= \leftarrow A_r$ a goal with $A_r$ an atom,  
and $R$ the depth-first, left-most control strategy. 
The {\em SLDNF$^*$-tree} $T_{N_r:G_r}$
for $P \cup \{G_r\}$ via $R$ is defined inductively as follows.

\begin{enumerate}
\item
$N_r:G_r$ is its root node, and the tree 
is completed once a 
node marked as {\bf LAST} is generated
or when all its leaf nodes have been marked as
$\Box_t$, $\Box_f$ or $\Box_{fl}$. 

\item 
\label{tree2} 
For each node $N_i: \leftarrow L_1,...,L_m$ 
in the tree that is selected by $R$, if $m=0$ then 
(1) $N_i$ is a {\em success} leaf marked as $\Box_t$
and (2) if $AL_{A_r@N_r}\neq \emptyset$ then
$N_i$ is also a node marked as {\bf LAST}. 
Otherwise (i.e. $m>0$), we distinguish between the following three cases. 
 
\begin{enumerate} 
\item 
\label{tree21} 
$L_1$ is a positive literal. For each clause 
$B \leftarrow B_1,...,B_n$ in $P$ such that $L_1$ and $B$  
are unifiable, $N_i$ has a child node 
 
$\qquad N_s: \leftarrow (B_1,...,B_n,L_2,...,L_m)\theta$ 
 
\noindent where $\theta$ is an mgu (i.e. most general unifier) of $L_1$ 
and $B$, the ancestor list for each $B_k\theta$, $k\in \{1, \ldots, n\}$, is  
$AL_{B_k\theta @N_s}=\{(N_i,L_1)\}\cup AL_{L_1@N_i}$, 
and the ancestor list for each $L_k\theta$, $k\in \{2, \ldots, m\}$,  
is $AL_{L_k\theta @N_s}=AL_{L_k@N_i}$. 
If there exists no clause in $P$ whose head can unify with $L_1$ then 
$N_i$ has a single child  node $-$ a 
{\em failure} leaf marked as $\Box_f$.  
  
\item 
\label{tree22} 
$L_1=\neg A$ is a ground negative literal. Let $T_{N_{i+1}:\leftarrow A}$ 
be an (subsidiary) SLDNF$^*$-tree for $P\cup \{\leftarrow A\}$ 
via $R$ with $AL_{A@N_{i+1}}=AL_{L_1@N_i}$. We have the following four 
cases:
\begin{enumerate}
\item
\label{tree221} 
$T_{N_{i+1}:\leftarrow A}$ has 
a success leaf. Then $N_i$ has a single child node  
$-$ a failure leaf marked as $\Box_f$.

\item
\label{tree222} 
$T_{N_{i+1}:\leftarrow A}$ has no success leaf but has 
a flounder leaf. Then $N_i$ has a single child node 
$-$ a flounder leaf marked as $\Box_{fl}$.
 
\item
\label{tree223} 
All branches of 
$T_{N_{i+1}:\leftarrow A}$ end with a failure leaf. Then  
$N_i$ has a single child node  
 
$\qquad N_s:\leftarrow L_2,...,L_m$ 
 
\noindent with $AL_{L_k@N_s}=AL_{L_k@N_i}$ for each 
$k\in \{2, \ldots, m\}$.

\item
\label{tree224}
Otherwise, $N_i$ has no child node. It is
the last node of $T_{N_r:G_r}$ so that it 
is marked as {\bf LAST}. 

\end{enumerate} 

\item
\label{tree23}
$L_1=\neg A$ is a non-ground negative literal. Then 
$N_i$ has a single child node
$-$ a {\em flounder} leaf marked as $\Box_{fl}$.
 
\end{enumerate} 
  
\end{enumerate} 
\end{definition}

Starting from the root node $N_r:G_r$,
we expand the nodes of the SLDNF$^*$-tree $T_{N_r:G_r}$ following
the depth-first order. The expansion for $T_{N_r:G_r}$ stops when either
a node marked as {\bf LAST} is generated or 
all of its leaf nodes have been marked
as $\Box_t$, $\Box_f$ or $\Box_{fl}$. 
 
In this paper we do not consider floundering 
$-$ a situation where 
a non-ground negative subgoal is selected by $R$ (see the case 
\ref{tree23}).
See \cite{chan88} for discussion on such topic. 

We first prove the following.

\begin{theorem}
\label{sontree}
Let $T_{N_{i+1}:\leftarrow A}$ be a subsidiary  SLDNF$^*$-tree 
built for proving a negative subgoal $L_1=\neg A$ at a node $N_i$
(see the case \ref{tree22}). 
Then $AL_{A@N_{i+1}}\neq \emptyset$.
\end{theorem}

\begin{proof}
Note that $AL_{A@N_{i+1}}=AL_{L_1@N_i}$. Since the subgoal 
$L_1$ at $N_i$ is negative, $N_i$ cannot be the root node of
the SLDNF$^*$-tree that contains $N_i$. Therefore, $L_1$ at $N_i$
has at least one ancestor subgoal (i.e. the subgoal at the root
node of the tree), which means $AL_{A@N_{i+1}}\neq \emptyset$.
\end{proof} 

In order to solve a top goal $G_0=\leftarrow A_0$, we build an
SLDNF$^*$-tree $T_{N_0:\leftarrow A_0}$ for $P\cup \{G_0\}$
via $R$ with $AL_{A_0@N_0} = \emptyset$. 
It is easy to see that $T_{N_0:\leftarrow A_0}$ is
an enhancement of the standard 
SLDNF-tree for $P\cup \{G_0\}$ via $R$ with 
the following three new features. 
\begin{enumerate} 
\item 
Each node $N_i$ is associated with an ancestor list $AL_{L_j@N_i}$ 
for each $L_j$ of its subgoals. 
In particular, subgoals 
of a subsidiary SLDNF$^*$-tree $T_{N_{i+1}:\leftarrow A}$ 
built for solving a negative subgoal 
$L_1=\neg A$ at $N_i$ inherit the ancestor list $AL_{L_1@N_i}$ 
(see the case \ref{tree22}). This bridges the ancestor-descendant
relationships across SLDNF$^*$-trees and is especially 
useful in identifying infinite derivations across 
SLDNF$^*$-trees (see Example \ref{eg1}). Note that a negative subgoal 
will never be an ancestor subgoal. 
 
\item 
In a standard SLDNF-tree, to handle a ground negative subgoal 
$L_1=\neg A$ at $N_i$ a full subsidiary 
SLDNF-tree $FT$ for $P\cup \{\leftarrow A\}$ via
$R$ must be generated. In an SLDNF$^*$-tree, however, 
the subsidiary SLDNF$^*$-tree 
$T_{N_{i+1}:\leftarrow A}$ may not include all branches of $FT$ 
because it will terminate at the first success leaf 
(see the case \ref{tree2} where by Theorem \ref{sontree}
$AL_{A@N_{i+1}}\neq \emptyset$).  
The intuition behind this is that it is absolutely 
unnecessary to exhaust the remaining branches of $FT$  
because they would never generate any new answers for $A$ (and $\neg A$). 
Such a pruning mechanism embedded in SLDNF$^*$-trees is very
useful in not only improving 
the efficiency of query evaluation but also avoiding some possible 
infinite derivations (see Example \ref{eg2}). In fact, Prolog   
performs the same pruning by using a {\em cut} 
operator to skip the remaining branches 
of $FT$ once the first success leaf is generated 
(e.g. see SICStus Prolog \cite{SICSTUS}).

\item
A well-known problem with the standard SLDNF-tree approach
(formally called {\em SLDNF-resolution} \cite{Cl79,Ld87}) is that for some
programs, such as $P=\{A\leftarrow \neg A\}$ and $G_0=\leftarrow A$, 
no SLDNF-trees exist \cite{AD94,KK89,MT92}. The main reason for
this abnormality lies in the fact that to solve a negative subgoal $\neg A$
it generates a subsidiary SLDNF-tree $FT$ for 
$P\cup \{\leftarrow A\}$ via $R$ {\em which is supposed either to contain
a success leaf or to consist of failure leaves}.
When $FT$ neither contains a success leaf nor finitely fails by
going into an infinite derivation,
the negative subgoal cannot be handled.  

In contrast, SLDNF$^*$-trees exist for any general logic programs.
A ground negative subgoal $\neg A$ at a node $N_i$ succeeds
if all branches of the subsidiary SLDNF$^*$-tree $T_{N_{i+1}:\leftarrow A}$ 
end with a failure leaf (see the case \ref{tree223}), and
fails if $T_{N_{i+1}:\leftarrow A}$ has a success leaf
(see the case \ref{tree221}). Otherwise, the value of the subgoal
$\neg A$ is undetermined and thus $N_i$ is marked
as {\bf LAST}, showing that it is the last node of the underlying
SLDNF$^*$-tree which can be finitely generated 
(see the case \ref{tree224}).\footnote{This case occurs when either
$T_{N_{i+1}:\leftarrow A}$ or some of its descendant SLDNF$^*$-trees
is infinite, or $T_{N_{i+1}:\leftarrow A}$ has an infinite number of
descendant SLDNF$^*$-trees. Note that {\bf LAST} is used here only
for the purpose of formulating an SLDNF$^*$-tree $-$ 
showing that $N_i$ is the last node of the SLDNF$^*$-tree. 
In practical implementation
of SLDNF$^*$-trees, in such a case $N_i$ will never be marked by
{\bf LAST} since it requires an infinitely long time to
build $T_{N_{i+1}:\leftarrow A}$ together with all
of its descendant SLDNF$^*$-trees. However, the SLDNF$^*$-tree
is always completed at $N_i$, whether $N_i$ is marked by {\bf LAST} or not,
because (1) such a case occurs at most one time in an 
SLDNF$^*$-tree and (2) it always occurs at the last generated node $N_i$.} 
The tree is then completed here. 

\end{enumerate} 
 
For convenience, we use dotted edges ``$\cdot\cdot\cdot\triangleright''$ 
to connect parent and child (subsidiary) SLDNF$^*$-trees,  
so that infinite derivations across SLDNF$^*$-trees 
can be clearly identified. Formally, we have

\begin{definition}
Let $P$ be a general logic program, $G_0$ a top goal
and $R$ the depth-first, left-most control strategy.
Let $T_{N_0:G_0}$ be the SLDNF$^*$-tree for $P \cup \{G_0\}$ via $R$ 
with $AL_{A_0@N_0} = \emptyset$.
A {\em generalized SLDNF-tree} for $P\cup \{G_0\}$ via $R$, 
denoted $GT_{G_0}$, is rooted at $N_0:G_0$ and consists of $T_{N_0:G_0}$ 
along with all its descendant SLDNF$^*$-trees, where
parent and child SLDNF$^*$-trees are connected
via ``$\cdot\cdot\cdot\triangleright''$. 
\end{definition}
 
Therefore, a path of a generalized SLDNF-tree may come  
across several SLDNF$^*$-trees through dotted edges. Any such a path 
starting at the root node $N_0:G_0$ is called a {\em generalized  
SLDNF-derivation}. 
 
Thus, there may occur two types of edges in a generalized SLDNF-derivation, 
``$\stackrel{C}{\longrightarrow}$'' 
and $``\cdot\cdot\cdot\triangleright''$. For convenience, 
we use $``\Rightarrow''$ to refer to either of them. 
Moreover, for any node $N_i:G_i$ we use 
$L_i^1$ to refer to the selected (i.e. left-most) subgoal in $G_i$. 
 
\begin{example} 
\label{eg1} 
{\em  
Let $P_1$ be a general logic program and $G_0$ a top goal, given by 
\begin{tabbing} 
$\qquad$ \= $P_1:$ $\quad$ \= $p(X)\leftarrow \neg p(f(X))$. \`$C_{p_1}$\\ 
\>          $G_0:$ \> $\leftarrow p(a).$ 
\end{tabbing} 
The generalized SLDNF-tree $GT_{\leftarrow p(a)}$ 
for $P_1\cup \{G_0\}$ is shown in 
Figure \ref{fig1}, where $\infty$ represents an infinite extension.
Note that to expand the node $N_1$, we build a subsidiary SLDNF$^*$-tree
$T_{N_2:\leftarrow p(f(a))}$. Since $T_{N_2:\leftarrow p(f(a))}$
neither contains a success leaf nor finitely fails
(i.e. not all of its leaf nodes are marked as $\Box_f$), $N_1$
is the last node of $T_{N_0:\leftarrow p(a)}$, marked as {\bf LAST}.
We see that $GT_{\leftarrow p(a)}$ is infinite, 
although all of its SLDNF$^*$-trees are finite. 
 
\begin{figure}[htb]
\centering

\setlength{\unitlength}{3947sp}%
\begingroup\makeatletter\ifx\SetFigFont\undefined%
\gdef\SetFigFont#1#2#3#4#5{%
  \reset@font\fontsize{#1}{#2pt}%
  \fontfamily{#3}\fontseries{#4}\fontshape{#5}%
  \selectfont}%
\fi\endgroup%
\begin{picture}(3975,1716)(676,-1006)
\thinlines
\put(2251, 89){\vector( 1, 0){0}}
\multiput(1876, 89)(75.00000,0.00000){5}{\makebox(1.6667,11.6667){\SetFigFont{5}{6}{\rmdefault}{\mddefault}{\updefault}.}}
\put(2776, 14){\vector( 0,-1){300}}
\put(4051,-436){\vector( 1, 0){0}}
\multiput(3676,-436)(75.00000,0.00000){5}{\makebox(1.6667,11.6667){\SetFigFont{5}{6}{\rmdefault}{\mddefault}{\updefault}.}}
\put(4576,-511){\vector( 0,-1){300}}
\put(1201,539){\vector( 0,-1){300}}
\put(4501,-961){\makebox(0,0)[lb]{\smash{\SetFigFont{10}{12.0}{\rmdefault}{\mddefault}{\updefault}$\infty$}}}
\put(1276,389){\makebox(0,0)[lb]{\smash{\SetFigFont{8}{9.6}{\rmdefault}{\mddefault}{\updefault}$C_{p_1}$}}}
\put(751,614){\makebox(0,0)[lb]{\smash{\SetFigFont{9}{10.8}{\rmdefault}{\mddefault}{\updefault}$N_0$:  $\leftarrow p(a)$}}}
\put(2851,-136){\makebox(0,0)[lb]{\smash{\SetFigFont{8}{9.6}{\rmdefault}{\mddefault}{\updefault}$C_{p_1}$}}}
\put(4651,-661){\makebox(0,0)[lb]{\smash{\SetFigFont{8}{9.6}{\rmdefault}{\mddefault}{\updefault}$C_{p_1}$}}}
\put(4126,-436){\makebox(0,0)[lb]{\smash{\SetFigFont{9}{10.8}{\rmdefault}{\mddefault}{\updefault}$N_4$:  $\leftarrow p(f(f(a)))$}}}
\put(2251,-436){\makebox(0,0)[lb]{\smash{\SetFigFont{9}{10.8}{\rmdefault}{\mddefault}{\updefault}$N_3$:  $\leftarrow \neg p(f(f(a)))$ }}}
\put(676, 89){\makebox(0,0)[lb]{\smash{\SetFigFont{9}{10.8}{\rmdefault}{\mddefault}{\updefault}$N_1$:  $\leftarrow \neg p(f(a))$ }}}
\put(2326, 89){\makebox(0,0)[lb]{\smash{\SetFigFont{9}{10.8}{\rmdefault}{\mddefault}{\updefault}$N_2$:  $\leftarrow p(f(a))$}}}
\put(976,-136){\makebox(0,0)[lb]{\smash{\SetFigFont{7}{8.4}{\rmdefault}{\mddefault}{\updefault}{\bf LAST}}}}
\put(2551,-661){\makebox(0,0)[lb]{\smash{\SetFigFont{7}{8.4}{\rmdefault}{\mddefault}{\updefault}{\bf LAST}}}}
\end{picture}

\caption{The generalized SLDNF-tree $GT_{\leftarrow p(a)}$.}\label{fig1} 
\end{figure}
}  
\end{example} 
 
\begin{example} 
\label{eg2} 
{\em 
Consider the following general logic program and top goal. 
\begin{tabbing} 
$\qquad$ \= $P_2:$ $\quad$ \= $p\leftarrow \neg q$. \`$C_{p_1}$\\ 
\>\> $q$.       \`$C_{q_1}$ \\ 
\>\> $q \leftarrow q$.            \`$C_{q_2}$ \\ 
\>          $G_0:$ \> $\leftarrow p.$ 
\end{tabbing} 
The generalized SLDNF-tree 
$GT_{\leftarrow p}$ for $P_2\cup \{G_0\}$ is depicted in 
Figure \ref{fig2} (a). 
$GT_{\leftarrow p}$ consists of two SLDNF$^*$-trees,
$T_{N_0:\leftarrow p}$ and $T_{N_2:\leftarrow q}$, which
are constructed as follows. Initially, $T_{N_0:\leftarrow p}$
has only the root node $N_0:\leftarrow p$. Expanding the root
node against the clause $C_{p_1}$ leads to the child node
$N_1:\leftarrow \neg q$. We then build a subsidiary SLDNF$^*$-tree
$T_{N_2:\leftarrow q}$ for $P_2\cup \{\leftarrow q\}$ via the
depth-first, left-most control strategy, where the expansion
stops right after the node $N_3$ is marked as {\bf LAST}.
Since $T_{N_2:\leftarrow q}$ has a success leaf, $N_1$ gets
a failure child node $N_5$. $T_{N_0:\leftarrow p}$ 
is then completed.

For the purpose of comparison,  
the standard SLDNF-trees for $P_2\cup \{\leftarrow p\}$  
are shown in Figure \ref{fig2} (b). 
Note that Figure \ref{fig2} (a) is finite,  
whereas Figure \ref{fig2} (b) is not. 
 
\begin{figure}[h]
\centering

\setlength{\unitlength}{3947sp}%
\begingroup\makeatletter\ifx\SetFigFont\undefined%
\gdef\SetFigFont#1#2#3#4#5{%
  \reset@font\fontsize{#1}{#2pt}%
  \fontfamily{#3}\fontseries{#4}\fontshape{#5}%
  \selectfont}%
\fi\endgroup%
\begin{picture}(5337,1971)(2776,-1261)
\put(3901,-1261){\makebox(0,0)[lb]{\smash{\SetFigFont{10}{12.0}{\rmdefault}{\mddefault}{\updefault}(a)}}}
\thinlines
\put(3301,-436){\vector(-2,-3){150}}
\put(7501,539){\vector(-2,-3){150}}
\put(7651,539){\vector( 2,-3){150}}
\put(7801, 14){\vector(-2,-3){150}}
\put(7951, 14){\vector( 2,-3){150}}
\put(4201, 14){\vector( 2,-3){150}}
\put(3601,-436){\vector( 2,-3){150}}
\put(3601,-211){\vector(-2,-3){0}}
\multiput(3751, 14)(-37.50000,-56.25000){4}{\makebox(1.6667,11.6667){\SetFigFont{5}{6}{\rmdefault}{\mddefault}{\updefault}.}}
\put(6751,-1261){\makebox(0,0)[lb]{\smash{\SetFigFont{10}{12.0}{\rmdefault}{\mddefault}{\updefault}(b)}}}
\put(8101,-436){\makebox(0,0)[lb]{\smash{\SetFigFont{10}{12.0}{\rmdefault}{\mddefault}{\updefault}$\infty$}}}
\put(3976,539){\vector( 0,-1){300}}
\put(6151, 14){\vector( 0,-1){300}}
\put(4051,389){\makebox(0,0)[lb]{\smash{\SetFigFont{8}{9.6}{\rmdefault}{\mddefault}{\updefault}$C_{p_1}$}}}
\put(3601, 89){\makebox(0,0)[lb]{\smash{\SetFigFont{9}{10.8}{\rmdefault}{\mddefault}{\updefault}$N_1$:  $\leftarrow \neg q$ }}}
\put(6226,389){\makebox(0,0)[lb]{\smash{\SetFigFont{8}{9.6}{\rmdefault}{\mddefault}{\updefault}$C_{p_1}$}}}
\put(5776,614){\makebox(0,0)[lb]{\smash{\SetFigFont{9}{10.8}{\rmdefault}{\mddefault}{\updefault}$N_0$:  $\leftarrow p$}}}
\put(5776, 89){\makebox(0,0)[lb]{\smash{\SetFigFont{9}{10.8}{\rmdefault}{\mddefault}{\updefault}$N_1$:  $\leftarrow \neg q$ }}}
\put(3601,614){\makebox(0,0)[lb]{\smash{\SetFigFont{9}{10.8}{\rmdefault}{\mddefault}{\updefault}$N_0$:  $\leftarrow p$}}}
\put(7126,389){\makebox(0,0)[lb]{\smash{\SetFigFont{8}{9.6}{\rmdefault}{\mddefault}{\updefault}$C_{q_1}$}}}
\put(7576, 89){\makebox(0,0)[lb]{\smash{\SetFigFont{9}{10.8}{\rmdefault}{\mddefault}{\updefault}$N_4$:  $\leftarrow q$}}}
\put(7801,389){\makebox(0,0)[lb]{\smash{\SetFigFont{8}{9.6}{\rmdefault}{\mddefault}{\updefault}$C_{q_2}$}}}
\put(7276,-436){\makebox(0,0)[lb]{\smash{\SetFigFont{9}{10.8}{\rmdefault}{\mddefault}{\updefault}$N_5$:  $\Box_t$}}}
\put(7426,-136){\makebox(0,0)[lb]{\smash{\SetFigFont{8}{9.6}{\rmdefault}{\mddefault}{\updefault}$C_{q_1}$}}}
\put(8101,-136){\makebox(0,0)[lb]{\smash{\SetFigFont{8}{9.6}{\rmdefault}{\mddefault}{\updefault}$C_{q_2}$}}}
\put(7276,614){\makebox(0,0)[lb]{\smash{\SetFigFont{9}{10.8}{\rmdefault}{\mddefault}{\updefault}$N_2$:  $\leftarrow q$}}}
\put(4126,-361){\makebox(0,0)[lb]{\smash{\SetFigFont{9}{10.8}{\rmdefault}{\mddefault}{\updefault}$N_5:\Box_f$}}}
\put(5776,-436){\makebox(0,0)[lb]{\smash{\SetFigFont{9}{10.8}{\rmdefault}{\mddefault}{\updefault}$N_{\infty}$:  $\Box_f$}}}
\put(3226,-361){\makebox(0,0)[lb]{\smash{\SetFigFont{9}{10.8}{\rmdefault}{\mddefault}{\updefault}$N_2$:  $\leftarrow q$}}}
\put(2926,-511){\makebox(0,0)[lb]{\smash{\SetFigFont{8}{9.6}{\rmdefault}{\mddefault}{\updefault}$C_{q_1}$}}}
\put(3676,-511){\makebox(0,0)[lb]{\smash{\SetFigFont{8}{9.6}{\rmdefault}{\mddefault}{\updefault}$C_{q_2}$}}}
\put(3526,-811){\makebox(0,0)[lb]{\smash{\SetFigFont{9}{10.8}{\rmdefault}{\mddefault}{\updefault}$N_4$:  $\leftarrow q$}}}
\put(2776,-811){\makebox(0,0)[lb]{\smash{\SetFigFont{9}{10.8}{\rmdefault}{\mddefault}{\updefault}$N_3$:  $\Box_t$}}}
\put(2851,-1036){\makebox(0,0)[lb]{\smash{\SetFigFont{7}{8.4}{\rmdefault}{\mddefault}{\updefault}{\bf LAST}}}}
\put(6976, 89){\makebox(0,0)[lb]{\smash{\SetFigFont{9}{10.8}{\rmdefault}{\mddefault}{\updefault}$N_3$:  $\Box_t$}}}
\put(6151,539){\vector( 0,-1){300}}
\end{picture}

\caption{The generalized SLDNF-tree $GT_{\leftarrow p}$ (a) 
and its two corresponding standard SLDNF-trees (b).}\label{fig2} 
\end{figure} 
} 
\end{example}

\section{Characterizing an Infinite Generalized SLDNF-Derivation} 
 
In this section we establish a necessary and sufficient condition for 
an infinite generalized SLDNF-derivation. We begin by introducing 
a few concepts.  
 
\begin{definition} 
\label{symbol-seq}  
Let $T$ be a term or an atom and $S$ be  
a string that consists of all predicate symbols, function  
symbols, constants 
and variables in $T$, which is obtained 
by reading these symbols sequentially from left to  
right. The {\em symbol string} of $T$, denoted 
$S_T$, is the string $S$ with every variable  
replaced by ${\cal X}$.  
\end{definition} 
 
For instance, let $T_1=a$, 
$T_2=f(X,g(X,f(a,Y)))$ and $T_3=[X,a]$. 
Then $S_{T_1}=a$, $S_{T_2}=f{\cal X}g{\cal X}fa{\cal X}$  
and $S_{T_3}=[{\cal X}|[a|[]]]$. Note that  
$[X,a]$ is a simplified representation for the 
list $[X|[a|[]]]$. 
 
\begin{definition} 
\label{sub-seq} 
Let $S_{T_1}$ and $S_{T_2}$ be two symbol strings. 
$S_{T_1}$ is a {\em projection} of $S_{T_2}$, denoted  
$S_{T_1}\subseteq_{proj}S_{T_2}$,  
if $S_{T_1}$ is obtained from $S_{T_2}$ by
removing zero or more elements. 
\end{definition} 
 
For example,  
$a{\cal X}{\cal X}bc\subseteq_{proj}fa{\cal X}e{\cal X}b{\cal X}cd$.  
It is easy to see 
that the relation $\subseteq_{proj}$ is reflexive and transitive. That is, 
for any symbol strings $S_1$, $S_2$ and $S_3$,  
we have $S_1\subseteq_{proj}S_1$, 
and that $S_1\subseteq_{proj}S_2$ and $S_2\subseteq_{proj}S_3$ implies 
$S_1\subseteq_{proj}S_3$. 
 
\begin{definition} 
\label{gvar} 
Let $A_1=p(.)$ and $A_2=p(.)$  
be two atoms. $A_1$ is said to {\em loop into} $A_2$, 
denoted $A_1\leadsto_{loop}A_2$, if  
$S_{A_1}\subseteq_{proj}S_{A_2}$. 
Let $N_i:G_i$ and $N_j:G_j$ be two nodes in a generalized 
SLDNF-derivation with $L_i^1\prec_{anc}L_j^1$ and 
$L_i^1 \leadsto_{loop} L_j^1$. 
Then $G_j$ is called a {\em loop goal} of $G_i$. 
\end{definition} 
 
The following result is immediate. 
\begin{theorem} 
\label{th} 
\begin{enumerate} 
\item 
The relation $\leadsto_{loop}$ is reflexive and transitive. 
\item 
If $A_1 \leadsto_{loop} A_2$ then $|A_1|\leq |A_2|$. 
\item 
If $G_3$ is a loop goal of $G_2$ that is a loop goal 
of $G_1$ then $G_3$ is a loop goal of $G_1$. 
\end{enumerate} 
\end{theorem} 
 
Observe that since a logic program has only
a finite number of clauses, an infinite generalized SLDNF-derivation results 
from repeatedly applying the same set of clauses, which leads to 
infinite repetition of selected variant subgoals or  
infinite repetition of selected subgoals with recursive 
increase in term size. By recursive increase of term size 
of a subgoal $A$ from a subgoal $B$ we mean that $A$ is $B$ 
with a few function/constant/variable symbols added  
and possibly with some variables changed to 
different variables. Such crucial dynamic 
characteristics of an infinite generalized SLDNF-derivation  
are captured by loop goals, as shown by the following principal  
theorem. 

\begin{theorem} 
\label{th4} 
$D$ is an infinite generalized SLDNF-derivation  
if and only if it is of the form 
\begin{tabbing} 
$\qquad$ $N_0:G_0\Rightarrow ... N_{g_1}:G_{g_1}\Rightarrow   
...N_{g_2}:G_{g_2}\Rightarrow ... N_{g_3}:G_{g_3}\Rightarrow ...$ 
\end{tabbing} 
such that for any $j \geq 1$, 
$G_{g_{j+1}}$ is a loop goal of $G_{g_j}$. 
\end{theorem}

We need Higman's Lemma to
prove this theorem.\footnote{It is one of the anonymous
reviewers who brought this lemma to our attention.}  
 
\begin{lemma}[(Higman's Lemma \cite{Hig52,Bol91})]
\label{hig} 
Let  $\{A_i\}_{i=1}^\infty$  be an infinite 
sequence of strings over a finite alphabet $\Sigma$.
Then for some  $i$  and  $k>i$,  $A_i \subseteq_{proj} A_k$
\end{lemma}

The following result follows from Lemma \ref{hig}.
 
\begin{lemma} 
\label{hig2}  
Let  $\{A_i\}_{i=1}^\infty$  be an infinite 
sequence of strings over a finite alphabet $\Sigma$.
Then there is an infinite 
increasing integer sequence $\{n_i\}_{i=1}^\infty$ such that for all $i$
$A_{n_i} \subseteq_{proj} A_{n_{i+1}}$.
\end{lemma} 
 
\begin{proof}\footnote{This proof is suggested by an anonymous
reviewer.}  
Suppose this is not true. Let us take a finite maximal subsequence
\[A_{n_1} \subseteq_{proj} A_{n_2} \subseteq_{proj} ... 
\subseteq_{proj} A_{n_{k_1}}\]
The subsequence is maximal in the sense that 
for no  $i > n_{k_1}$  do we have 
$A_{n_{k_1}} \subseteq_{proj} A_i$. 
We know that such a subsequence with length 
at least 2 must exist from Lemma \ref{hig} and the assumption that the 
assertion of the lemma does not hold for the sequence $\{A_i\}_{i=1}^\infty$.
Now look at the elements of the original sequence with indices larger
than $n_{k_1}$ and take another such finite maximal subsequence from them. 
Continuing in this way,
we get infinitely many such maximal subsequences. 
Let  $\{A_{n_{k_i}}\}_{i=1}^\infty$ be the sequence 
of last elements of the maximal subsequences. 
By Lemma \ref{hig}, this sequence has two elements,
$A_{n_{k_i}}$ and $A_{n_{k_j}}$ with $n_{k_i} < n_{k_j}$,
such that $A_{n_{k_i}} \subseteq_{proj} A_{n_{k_j}}$. 
This contradicts the assumption that $A_{n_{k_i}}$ is
the last element of some finite maximal subsequence. \end{proof}

The following lemma is needed to prove Theorem \ref{th4}.

\begin{lemma} 
\label{th2} 
Let $D$ be an infinite generalized SLDNF-derivation.  
Then there are infinitely many goals 
$G_{g_1}, G_{g_2}, ...$  
in $D$ such that for any $j \geq 1$, 
$L_j^1 \prec_{anc} L_{j+1}^1$.
\end{lemma} 
 
\begin{proof} 
Let $D$  be of the form 
\[N_0:G_0\Rightarrow N_1:G_1 \Rightarrow ... \Rightarrow 
N_i:G_i\Rightarrow N_{i+1}:G_{i+1} \Rightarrow ...\]
Consider derivation steps like  
$N_i:G_i \stackrel{C}{\longrightarrow} N_{i+1}:G_{i+1}
 \cdot\cdot\cdot\triangleright N_{i+2}:G_{i+2}$, where  
$L_i^1$ is a positive subgoal and $L_{i+1}^1=\neg A$ a negative subgoal.
So $L_{i+2}^1=A$. We see that both $L_i^1$ and $L_{i+1}^1$ need the proof of 
$L_{i+2}^1$. Moreover, given $L_{i+2}^1$ the provability of 
$L_i^1$ does not depend on
$L_{i+1}^1$. Since $L_{i+1}^1$  has no descendant subgoals, removing
it would affect neither the provability nor the ancestor-descendant
relationships of other subgoals in $D$. Therefore, we delete  
$L_{i+1}^1$ by marking $N_{i+1}$ with \#.

For each derivation step  
$N_i:G_i \stackrel{C}{\longrightarrow} N_{i+1}:G_{i+1}$, where  
$L_i^1$ is a positive subgoal and $C=A\leftarrow B_1,...B_n$ 
such that $A\theta = L_i^1\theta$ under an mgu $\theta$, we do the following:  
\begin{enumerate} 
\item 
If $n=0$, which means $L_i^1$ is proved at this step, 
mark node $N_i$ with \#. 
 
\item 
Otherwise, the proof of $L_i^1$ needs the proof of $B_j\theta$ $(j=1,...,n)$. 
If all descendant nodes of $N_i$ in $D$ have been marked with \#, which means that  
all $B_j\theta$ have been proved at some steps in $D$, mark node $N_i$ 
with \#. 
\end{enumerate} 
 
Note that the root node $N_0$ will never be marked with \#, for otherwise 
$G_0$ would have been proved and $D$ should have ended at a success 
or failure leaf. After the  
above marking process, let $D$ become 
\begin{tabbing} 
$\qquad$ $N_0:G_0\Rightarrow ... \Rightarrow N_{i_1}:G_{i_1} \Rightarrow ... \Rightarrow 
N_{i_2}:G_{i_2}\Rightarrow ... \Rightarrow  N_{i_k}:G_{i_k} \Rightarrow ...$ 
\end{tabbing} 
where all nodes except $N_0,N_{i_1},N_{i_2},...,N_{i_k},...$ are marked with 
\#. Since we use the depth-first, left-most control strategy, for any  
$j\geq 0$ the proof of $L_{i_j}^1$ needs the proof of $L_{i_{j+1}}^1$ (let 
$i_0=0$), for otherwise $N_{i_j}$ would have been marked with \#. That is, 
$L_{i_j}^1$ is an ancestor subgoal of $L_{i_{j+1}}^1$.  
Moreover, $D$ must contain  
an infinite number of such nodes because  
if $N_{i_k}:G_{i_k}$ was the last one, 
which means that all nodes after $N_{i_k}$ were  
marked with \#, then $L_{i_k}^1$ 
would be proved, so that $N_{i_k}$ should be marked with \#, a contradiction. 
\end{proof}
 
We are ready to prove Theorem \ref{th4}. 
 
\begin{proof}[(Proof of Theorem \ref{th4})]
$(\Longleftarrow)$ Straightforward. 
 
$(\Longrightarrow)$  
By Lemma \ref{th2}, $D$ contains an infinite 
sequence of selected subgoals $H_1=\{L_{j_i}^1\}_{i=1}^\infty$ such that  
for any $i$ $L_{j_i}^1 \prec_{anc} L_{j_{i+1}}^1$.
Since any logic program has only a finite number of predicate  
symbols, $H_1$ must have 
an infinite subsequence $H_2=\{L_{k_i}^1\}_{i=1}^\infty$ 
such that all $L_{k_i}^1$ have the same predicate symbol, say $p$. 
We now show that  
$H_2$ has an infinite subsequence  
$\{L_{g_i}^1\}_{i=1}^\infty$ 
such that for any $i$ $L_{g_i}^1\leadsto_{loop}L_{g_{i+1}}^1$. 
 
Let $T$ be the (finite) set of all constant 
and function symbols in the logic program and let $\Sigma = T\cup \{{\cal X}\}$.  
Then the symbol string  
$S_{L_{k_i}^1}$ of each $L_{k_i}^1$ in $H_2$ is a string over $\Sigma$ 
that begins with $p$. These symbol strings 
constitute an infinite sequence $\{p A_i\}_{i=1}^\infty$
with each $A_i$ being a substring.
By Lemma \ref{hig2} there is an infinite increasing 
integer sequence $\{n_i\}_{i=1}^\infty$ such that
for any $i$ $p A_{n_i} \subseteq_{proj} p A_{n_{i+1}}$. 
Therefore, $H_2$ has an infinite subsequence  
$H_3=\{L_{g_i}^1\}_{i=1}^\infty$ 
with $S_{L_{g_i}^1}=p A_{n_i}$ 
being the symbol string of $L_{g_i}^1$. That is, for any $i$ 
$S_{L_{g_i}^1}\subseteq_{proj}S_{L_{g_{i+1}}^1}$. 
Thus, for any $i$ $L_{g_i}^1\leadsto_{loop}L_{g_{i+1}}^1$. 
\end{proof} 
 
\section{Characterizing Termination of General Logic Programs} 
\label{verify}

In \cite{DD93}, a generic definition of termination of 
logic programs is presented as follows. 
 
\begin{definition}[(\cite{DD93})] 
\label{term-def1} 
Let $P$ be a general logic program, $S_Q$ a set of queries and $S_R$ 
a set of selection rules. $P$ is terminating w.r.t. $S_Q$ and $S_R$ 
if for each query $Q_i$ in $S_Q$ and for each selection rule $R_j$ in $S_R$,  
all SLDNF-trees for $P\cup \{\leftarrow Q_i\}$ via $R_j$ are finite.  
\end{definition} 
 
Observe that the above definition considers finite SLDNF-trees  
for termination. That is, $P$ is terminating w.r.t. $Q_i$ only if all (complete) 
SLDNF-trees for $P\cup \{\leftarrow Q_i\}$ are finite. This does not 
seem to apply to Prolog where there exist cases in which $P$ is terminating 
w.r.t. $Q_i$ and $R_j$, although some (complete) SLDNF-trees  
for $P\cup \{\leftarrow Q_i\}$ are 
infinite. Example \ref{eg2} gives such an illustration, where Prolog 
terminates with a negative answer to the top goal $G_0$. 
 
In view of the above observation, we present the following  
slightly different definition of termination based on  
a generalized SLDNF-tree. 
 
\begin{definition} 
\label{term-def2} 
Let $P$ be a general logic program, $S_Q$ a finite set of queries and $R$ 
the depth-first, left-most control strategy.  
$P$ is terminating w.r.t. $S_Q$ and $R$ 
if for each query $Q_i$ in $S_Q$, the generalized SLDNF-tree 
for $P\cup \{\leftarrow Q_i\}$ via $R$ is finite.  
\end{definition} 
 
The above definition implies that $P$ is terminating w.r.t. $S_Q$ and $R$  
if and only if there is no infinite generalized SLDNF-derivation 
in any generalized SLDNF-tree $GT_{\leftarrow Q_i}$.  
This obviously applies to Prolog.  
We then have the following immediate result from Theorem \ref{th4},
which characterizes termination of a general logic program. 
 
\begin{theorem} 
\label{th-iff} 
$P$ is terminating w.r.t. $S_Q$ and $R$ if and only if  
for each query $Q_i$ in $S_Q$ 
there is no generalized SLDNF-derivation  
in $GT_{\leftarrow Q_i}$ of the form 
\begin{tabbing} 
$\qquad$ $N_0:G_0\Rightarrow ... N_{g_1}:G_{g_1}\Rightarrow   
...N_{g_2}:G_{g_2}\Rightarrow ... N_{g_3}:G_{g_3}\Rightarrow ...$ 
\end{tabbing} 
such that for any $j \geq 1$,  
$G_{g_{j+1}}$ is a loop goal of $G_{g_j}$. 
\end{theorem} 

\section{Related Work} 
\label{related-work}

Concerning termination analysis, we refer the reader to the papers of 
Decorte, De Schreye and Vandecasteele \cite{DD93,DDV99}  
for a comprehensive bibliography. 
Most existing termination analysis techniques are 
static approaches, which only make use of the syntactic structure of 
the source code of a logic program to establish some well-founded 
conditions/constraints that, when satisfied, yield a termination  
proof. Since non-termination is caused by an infinite generalized 
SLDNF-derivation, which contains some essential dynamic characteristics 
that are hard to capture in a static way, 
static approaches appear to be less precise than a dynamic one. 
For example, it is difficult to apply a static approach to prove 
the termination of program $P_2$ in Example \ref{eg2} with respect to 
a query pattern $p$.

The concept of generalized SLDNF-trees is the basis of our approach.
There are several new definitions of SLDNF-trees
presented in the literature, such as that of
Apt and Doets \cite{AD94}, Kunen \cite{KK89},
or Martelli and Tricomi \cite{MT92}. 
Generalized SLDNF-trees have two distinct features as 
compared to these definitions.
First, the ancester-descendent relation is explicitly expressed
(using ancester lists) in a generalized SLDNF-tree, which
is essential in identifying loop goals.
Second, a ground negative subgoal $\neg A$ at a node $N_i$ in a 
SLDNF$^*$-tree $T_{N_r:G_r}$ is formulated in the same way
as in Prolog, i.e. (1) the subsidiary SLDNF$^*$-tree 
$T_{N_{i+1}:\leftarrow A}$ for the 
subgoal terminates at the first success leaf, and
(2) $\neg A$ succeeds
if all branches of $T_{N_{i+1}:\leftarrow A}$ 
end with a failure leaf and
fails if $T_{N_{i+1}:\leftarrow A}$ has a success leaf.
When $T_{N_{i+1}:\leftarrow A}$ goes into an infinite
extension, the node $N_i$ is treated as the last node of 
$T_{N_r:G_r}$ which can be finitely generated.
As a result, a generalized SLDNF-tree
exists for any general logic programs.

Our work is also related to loop checking $-$ another research topic  
in logic programming which focuses 
on detecting and eliminating infinite loops.
Informally, a derivation 
\begin{center} 
$N_0:G_0 \Rightarrow N_1:G_1 \Rightarrow ...\Rightarrow 
N_i:G_i \Rightarrow ... 
\Rightarrow N_k:G_k \Rightarrow ...$ 
\end{center} 
is said to step into a loop at a node $N_k:G_k$ if there is a 
node $N_i:G_i$ ($0 \leq i < k$) in the 
derivation such that $G_i$ and $G_k$ are {\em sufficiently 
similar}. Many mechanisms related to loop checking  
have been presented in the literature 
(e.g. \cite{BAK91,shen001}). 
However, most of them apply only to {\em SLD-derivations}
for positive logic programs and thus 
cannot deal with infinite recursions through  
negation like that in Figure \ref{fig1} or \ref{fig2}.

Loop goals are defined on a generalized SLDNF-derivation for general
logic programs and can be used to define the sufficiently similar
goals in loop checking. For such an application, they play a role
similar to {\em expanded variants} defined in \cite{shen001}. 
Informally, expanded variants are variants except that some terms  
may grow bigger. However, expanded variants 
have at least three disadvantages as compared to loop goals:
their definition is less intuitive, their computation
is more expensive, and they
are not transitive in the sense that $A$ being an expanded variant of 
$B$ that is an expanded variant of $C$ does not necessarily 
imply $A$ is an expanded variant of $C$.  
 
\section{Conclusions and Future Work}

We have presented an approach to characterizing termination of 
general logic programs by making use of dynamic features. 
A concept of generalized SLDNF-trees
is introduced, a necessary and sufficient condition 
for infinite generalized SLDNF-derivations 
is established, and a new characterization
of termination of a general logic program is developed.

We have recently developed an algorithm for automatically
predicting termination of general logic programs based on the
characterization established in this paper. The algorithm 
identifies the most-likely non-terminating programs. 
Let $P$ be a general logic program, 
$S_Q$ a set of queries and $R$ the depth-first, left-most control strategy.
$P$ is said to be {\em most-likely} non-terminating w.r.t. $S_Q$ and $R$ 
if for some query $Q_i$ in $S_Q$, there is a generalized SLDNF-derivation
with a few (e.g. two or three) loop goals.
Our experiments show that for most representative
general logic programs we have collected in the literature, 
they are not terminating  w.r.t. $S_Q$ and $R$ if and only if
they are most-likely non-terminating w.r.t. $S_Q$ and $R$.
This algorithm can be incorporated into Prolog as a debugging tool,
which would provide the users with valuable debugging 
information for them to understand the causes of non-termination.  
 
Tabled logic programming is receiving increasing attention  
in the community of logic programming 
(e.g. \cite{chen96,shen003}).  
Verbaeten, De Schreye and Sagonas \cite{VDK001}  
recently exploited termination proofs for positive logic programs 
with tabling. For future research, we are considering extending the work of  
the current paper to deal with general logic programs with tabling.
 
\begin{acks} 
We are particularly grateful to Danny De Schreye, 
Krzysztof Apt, Jan Willem Klop and  
the three anonymous referees for their  
constructive comments on our work 
and valuable suggestions for its improvement.
One anonymous reviewer brought the Higman's lemma to our attention.
Another anonymous reviewer suggested the proof of Lemma \ref{hig2}
and presented us a different yet interesting proof
to the Higman's lemma based on an idea from Dershowitz's paper \cite{Der87}.
The work of Yi-Dong Shen is supported in part by Chinese National 
Natural Science Foundation and Trans-Century Training Program 
Foundation for the Talents by the Chinese Ministry of Education.
Qiang Yang is supported by NSERC and IRIS III grants.   
\end{acks}

\bibliographystyle{acmtrans}

\begin{thebibliography}{}

\bibitem[\protect\citeauthoryear{Apt and Doets}{Apt and Doets}{1994}]{AD94}
{\sc Apt, K.~R.} {\sc and} {\sc Doets, K.} 1994.
\newblock A new definition of sldnf-resolution.
\newblock {\em Journal of Logic Programming\/}~{\em 18}, 177--190.

\bibitem[\protect\citeauthoryear{Apt and Pedreschi}{Apt and
  Pedreschi}{1993}]{Apt1}
{\sc Apt, K.~R.} {\sc and} {\sc Pedreschi, D.} 1993.
\newblock Reasoning about termination of pure prolog programs.
\newblock {\em Information and Computation\/}~{\em 106}, 109--157.

\bibitem[\protect\citeauthoryear{Bezem}{Bezem}{1992}]{Bezem92}
{\sc Bezem, M.} 1992.
\newblock Characterizing termination of logic programs with level mapping.
\newblock {\em Journal of Logic Programming\/}~{\em 15,\/}~1/2, 79--98.

\bibitem[\protect\citeauthoryear{Bol}{Bol}{1991}]{Bol91}
{\sc Bol, R.~N.} 1991.
\newblock Loop checking in logic programming.
\newblock Ph.D. thesis, The University of Amsterdam, Amsterdam.

\bibitem[\protect\citeauthoryear{Bol, Apt, and Klop}{Bol
  et~al\mbox{.}}{1991}]{BAK91}
{\sc Bol, R.~N.}, {\sc Apt, K.~R.}, {\sc and} {\sc Klop, J.~W.} 1991.
\newblock An analysis of loop checking mechanisms for logic programs.
\newblock {\em Theoretical Computer Science\/}~{\em 86,\/}~1, 35--79.

\bibitem[\protect\citeauthoryear{Bossi, Cocco, and Fabris}{Bossi
  et~al\mbox{.}}{1994}]{BCF94}
{\sc Bossi, A.}, {\sc Cocco, N.}, {\sc and} {\sc Fabris, M.} 1994.
\newblock Norms on terms and their use in proving universal termination of a
  logic program.
\newblock {\em Theoretical Computer Science\/}~{\em 124,\/}~1, 297--328.

\bibitem[\protect\citeauthoryear{Brodsky and Sagiv}{Brodsky and
  Sagiv}{1991}]{BS91}
{\sc Brodsky, A.} {\sc and} {\sc Sagiv, Y.} 1991.
\newblock Inference of inequality constraints in logic programs.
\newblock In {\em Proceedings of the Tenth ACM SIGACT-SIGMOD-SIGART Symposium
  on Principles of Database Systems}. ACM Press, Denver, 227--240.

\bibitem[\protect\citeauthoryear{Chan}{Chan}{1988}]{chan88}
{\sc Chan, D.} 1988.
\newblock Constructive negation based on the completed database.
\newblock In {\em Proceedings of the Fifth International Conference and
  Symposium on Logic Programming}. MIT Press, Seattle, 111--125.

\bibitem[\protect\citeauthoryear{Chen and Warren}{Chen and
  Warren}{1996}]{chen96}
{\sc Chen, W.~D.} {\sc and} {\sc Warren, D.~S.} 1996.
\newblock Tabled evaluation with delaying for general logic programs.
\newblock {\em J. ACM\/}~{\em 43,\/}~1, 20--74.

\bibitem[\protect\citeauthoryear{Clark}{Clark}{1978}]{Cl79}
{\sc Clark, K.~L.} 1978.
\newblock Negation as failure.
\newblock In {\em (H.Gallaire and J. Minker, eds.) Logic and Databases}.
  Plenum, New York, 293--322.

\bibitem[\protect\citeauthoryear{Decorte, Schreye, and Fabris}{Decorte
  et~al\mbox{.}}{1993}]{DDF93}
{\sc Decorte, S.}, {\sc Schreye, D.~D.}, {\sc and} {\sc Fabris, M.} 1993.
\newblock Automatic inference of norms: A missing link in automatic termination
  analysis.
\newblock In {\em Proceedings of the 1993 International Symposium on Logic
  Programming}. MIT Press, Vancouver, Canada, 420--436.

\bibitem[\protect\citeauthoryear{Decorte, Schreye, and Vandecasteele}{Decorte
  et~al\mbox{.}}{1999}]{DDV99}
{\sc Decorte, S.}, {\sc Schreye, D.~D.}, {\sc and} {\sc Vandecasteele, H.}
  1999.
\newblock Constraint-based termination analysis of logic programs.
\newblock {\em ACM Transactions on Programming Languages and Systems\/}~{\em
  21,\/}~6, 1137--1195.

\bibitem[\protect\citeauthoryear{Dershowitz}{Dershowitz}{1987}]{Der87}
{\sc Dershowitz, N.} 1987.
\newblock Termination of rewriting.
\newblock {\em J. Symbolic Computation\/}~{\em 3}, 69--116.

\bibitem[\protect\citeauthoryear{Higman}{Higman}{1952}]{Hig52}
{\sc Higman, G.} 1952.
\newblock Ordering by divisibility in abstract algebras.
\newblock {\em Proceedings of the London Mathematical Society\/}~{\em 3,\/}~2,
  326--336.

\bibitem[\protect\citeauthoryear{ISLAB}{ISLAB}{1998}]{SICSTUS}
{\sc ISLAB}. 1998.
\newblock {\em SICStus Prolog User's Manual}.
\newblock Intelligent Systems Laboratory, Swedish Institute of Computer
  Science, Available from
  http://www.sics.se/sicstus/docs/3.7.1/html/sicstus\_toc.html.

\bibitem[\protect\citeauthoryear{Kunen}{Kunen}{1989}]{KK89}
{\sc Kunen, K.} 1989.
\newblock Signed data dependencies in logic programming.
\newblock {\em Journal of Logic Programming\/}~{\em 7}, 231--246.

\bibitem[\protect\citeauthoryear{Lindenstrauss and Sagiv}{Lindenstrauss and
  Sagiv}{1997}]{LS97}
{\sc Lindenstrauss, N.} {\sc and} {\sc Sagiv, Y.} 1997.
\newblock Automatic termination analysis of logic programs.
\newblock In {\em Proceedings of the Fourteenth International Conference on
  Logic Programming}. MIT Press, Leuven, Belgium, 63--77.

\bibitem[\protect\citeauthoryear{Lloyd}{Lloyd}{1987}]{Ld87}
{\sc Lloyd, J.~W.} 1987.
\newblock {\em Foundations of Logic Programming}.
\newblock Springer-Verlag, Berlin.

\bibitem[\protect\citeauthoryear{Martelli and Tricomi}{Martelli and
  Tricomi}{1992}]{MT92}
{\sc Martelli, M.} {\sc and} {\sc Tricomi, C.} 1992.
\newblock A new sldnf-tree.
\newblock {\em Information Processing Letters\/}~{\em 43,\/}~2, 57--62.

\bibitem[\protect\citeauthoryear{Martin, King, and Soper}{Martin
  et~al\mbox{.}}{1997}]{MKS96}
{\sc Martin, J.~C.}, {\sc King, A.}, {\sc and} {\sc Soper, P.} 1997.
\newblock Typed norms for typed logic programs.
\newblock In {\em Proceedings of the 6th International Workshop on Logic
  Programming Synthesis and Transformation}. Springer, Stockholm, Sweden,
  224--238.

\bibitem[\protect\citeauthoryear{Plumer}{Plumer}{1990a}]{Pl90b}
{\sc Plumer, L.} 1990a.
\newblock {\em Termination Proofs for Logic Programs}.
\newblock Lecture Notes in Computer Science 446, Springer-Verlag, Berlin.

\bibitem[\protect\citeauthoryear{Plumer}{Plumer}{1990b}]{Pl90a}
{\sc Plumer, L.} 1990b.
\newblock Termination proofs for logic programs based on predicate
  inequalities.
\newblock In {\em Proceedings of the Seventh International Conference on Logic
  Programming}. MIT Press, Cambridge, MA, 634--648.

\bibitem[\protect\citeauthoryear{Schreye and Decorte}{Schreye and
  Decorte}{1993}]{DD93}
{\sc Schreye, D.~D.} {\sc and} {\sc Decorte, S.} 1993.
\newblock Termination of logic programs: the never-ending story.
\newblock {\em Journal of Logic Programming\/}~{\em 19,\/}~20, 199--260.

\bibitem[\protect\citeauthoryear{Schreye and Verschaetse}{Schreye and
  Verschaetse}{1995}]{DV95}
{\sc Schreye, D.~D.} {\sc and} {\sc Verschaetse, K.} 1995.
\newblock Deriving linear size relations for logic programs by abstract
  interpretation.
\newblock {\em New Generation Computing\/}~{\em 13,\/}~2, 117--154.

\bibitem[\protect\citeauthoryear{Schreye, Verschaetse, and Bruynooghe}{Schreye
  et~al\mbox{.}}{1992}]{DVB92}
{\sc Schreye, D.~D.}, {\sc Verschaetse, K.}, {\sc and} {\sc Bruynooghe, M.}
  1992.
\newblock A framework for analyzing the termination of definite logic programs
  with respect to call patterns.
\newblock In {\em Proceedings of the International Conference on Fifth
  Generation Computer Systems}. IOS Press, Tokyo, Japan, 481--488.

\bibitem[\protect\citeauthoryear{Shen, Yuan, and You}{Shen
  et~al\mbox{.}}{2001}]{shen001}
{\sc Shen, Y.~D.}, {\sc Yuan, L.~Y.}, {\sc and} {\sc You, J.~H.} 2001.
\newblock Loop checks for logic programs with functions.
\newblock {\em Theoretical Computer Science\/}~{\em 266,\/}~1/2, 441--461.

\bibitem[\protect\citeauthoryear{Shen, Yuan, and You}{Shen
  et~al\mbox{.}}{2002}]{shen003}
{\sc Shen, Y.~D.}, {\sc Yuan, L.~Y.}, {\sc and} {\sc You, J.~H.} 2002.
\newblock Slt-resolution for the well-founded semantics.
\newblock {\em Journal of Automated Reasoning\/}~{\em 28,\/}~1, 53--97.

\bibitem[\protect\citeauthoryear{Ullman and Gelder}{Ullman and
  Gelder}{1988}]{UVG88}
{\sc Ullman, J.~D.} {\sc and} {\sc Gelder, A.~V.} 1988.
\newblock Efficient tests for top-down termination of logical rules.
\newblock {\em J. ACM\/}~{\em 35,\/}~2, 345--373.

\bibitem[\protect\citeauthoryear{Verbaeten, Schreye, and Sagonas}{Verbaeten
  et~al\mbox{.}}{2001}]{VDK001}
{\sc Verbaeten, S.}, {\sc Schreye, D.~D.}, {\sc and} {\sc Sagonas, K.} 2001.
\newblock Termination proofs for logic programs with tabling.
\newblock {\em ACM Transactions on Computational Logic\/}~{\em 2,\/}~1, 57--92.

\bibitem[\protect\citeauthoryear{Verschaetse}{Verschaetse}{1992}]{V92}
{\sc Verschaetse, K.} 1992.
\newblock Static termination analysis for definite horn clause programs.
\newblock Ph.D. thesis, Department of Computer Science, K. U. Leuven, Available
  at http://www.cs.kuleuven.ac.be/~lpai.

\end{thebibliography}

\begin{received}
Received June 2000;
revised August 2001 and December 2001;
accepted April 2002
\end{received}

\end{document}